# The radio-frequency quadrupole

*Maurizio Vretenar*
CERN, Geneva, Switzerland

**Abstract**
Radio-frequency quadrupole (RFQ) linear accelerators appeared on the accelerator scene in the late 1970s and have since revolutionized the domain of low-energy proton and ion acceleration. The RFQ makes the reliable production of unprecedented ion beam intensities possible within a compact radio-frequency (RF) resonator which concentrates the three main functions of the low-energy linac section: focusing, bunching and accelerating. Its sophisticated electrode structure and strict beam dynamics and RF requirements, however, impose severe constraints on the mechanical and RF layout, making the construction of RFQs particularly challenging. This lecture will introduce the main beam optics, RF and mechanical features of a RFQ emphasizing how these three aspects are interrelated and how they contribute to the final performance of the RFQ.

## 1      The challenges of low-energy acceleration of hadron beams

The low-energy section, between the ion source and the first drift-tube-based accelerating structure, is probably the most complicated part of any hadron linear accelerator. It is in this part of the linac that the following conditions are met:

(a) Defocusing due to space charge forces (mutually repulsive Coulomb forces between beam particles) is the highest. The space charge force acting on a single particle is inversely proportional to $\gamma^2$ ($\gamma$ is the relativistic parameter here) and starts decreasing as soon as the beam becomes relativistic and the attraction between particles travelling close to the speed of light compensates for the Coulomb repulsion. The reduction will become perceptible only above few megaelectronvolts beam energy, however, leaving space charge at its maximum at energies below. To compensate for space charge, external focusing must be the highest in the low-energy range: in the usual approach, this means short focusing periods and a large number of high-gradient quadrupoles. A strong limitation to the focusing achievable at low energy, however, comes from the small dimensions of the accelerating cells. In a drift-tube structure, the distance between the centres of two quadrupoles placed inside drift tubes is $\beta\lambda$; considering that some length on the beam axis is taken by the gap and by the metal of the tubes, the space available for the quadrupole is only about $\beta\lambda/2$. At 1 MeV, $\beta$ = 4.6% and for $\lambda \sim$ 1 m the maximum length of a quadrupole is about 20 mm, nearly the same as the required aperture. The quadrupole would be dominated by fringe fields and it would be impossible to achieve on the axis a gradient sufficient to control high space charge forces.

(b) The continuous beam coming out of the source has to be bunched to be accelerated in the first radio-frequency (RF) accelerating structure. The process of bunching by means of longitudinally focusing RF forces is a critical operation: it defines the longitudinal beam emittance and can lead to the loss of a large fraction of the particles if the resulting emittance is not matched to the acceptance of the first accelerating structure.

(c) Usual low-energy accelerating structures have reduced RF efficiency and high mechanical complexity, because of the need to adapt the length of every cell to the beam velocity. Short cells have high stray capacitances that for a given power dissipation reduce the effective voltage available for the beam, or in other terms they are particularly ineffective in concentrating on the axis the electric field required for acceleration. The result is that the accelerator cost per meter (or per megaelectronvolt acceleration) in this section tends to be the highest of all of the parts of the linac and needs to be carefully optimized.

Before the invention of the radio-frequency quadrupole (RFQs), the classical solution to cover this critical energy range was to extend as much as possible the extraction voltage from the ion source and to start the first accelerating structure, usually a drift-tube linac (DTL), from the lowest possible energy. The use of large high-voltage (HV) generators, at the limit of technology, allowed extracting from the source a beam with sufficient velocity to be injected into a DTL of relatively low frequency (to increase $\lambda$) equipped with special short quadrupoles in the first drift tubes. In these systems, however, the maximum beam current was limited by the size and aperture of the first quadrupoles and by the space charge in the transport line between the source and the DTL. Moreover, the need for a low RF frequency reduced the overall acceleration efficiency.

In the old low-energy beam transport (LEBT) lines bunching was provided by a single-gap RF cavity followed by a drift space before the DTL. The RF cavity applies a small sinusoidal modulation to the energy and velocity of the beam; after the drift, particles that were on the rising slope of the modulating voltage tend to group together, the particles that arrived first in the cavity being slower and those arriving later being faster. This will result in a higher density of particles around the particle whose energy was not changed (on the rising part of the voltage), which will be maximum at a given distance from the RF cavity. If the first accelerating gap of the DTL is placed exactly in this position, a large fraction of the beam will lie within the acceptance ("bucket") of the DTL and will be accelerated, but another fraction will be outside and will be lost in the first gaps of the DTL. A single cavity bunching section has a low transmission, of the order of 50%, and requires long drift distances where space charge can easily lead to emittance growth. To increase transmission, many bunching systems included a second harmonic cavity after the first, to linearize the overall voltage seen by the beam and extend the capture region.

As an example of low-energy section before the RFQ, Figs. 1 and 2 show the Linac2 installation at CERN as it looked between construction (in 1976) and the installation of a new RFQ replacing the original injector (in 1993). At the time of construction this was one of the proton linacs with the highest beam current in the world, 150 mA. The beam was extracted from the ion source at 750 keV, a voltage produced by a large Cockcroft–Walton generator placed in a separate HV room (Fig. 1). The 750 keV line (Fig. 2) was 5.6 m long and included four quadrupole triplets, diagnostics equipment and a double-harmonic bunching system, made of a first single-gap cavity at 202 MHz frequency followed by another at 404 MHz. The trapping efficiency of this line (ratio between current accelerated in the DTL and current extracted from the source) was as high as 80%, thanks to a careful design and to the double-harmonic bunching [1].

In spite of its sophisticated design and construction, for all operating linacs the low-energy section was not only the bottleneck in terms of beam current because of space charge limitations and of low bunching efficiency, but was also one of the main limitations in terms of reliability, the HV required for the injector being the origin of a large fraction of machine downtime.

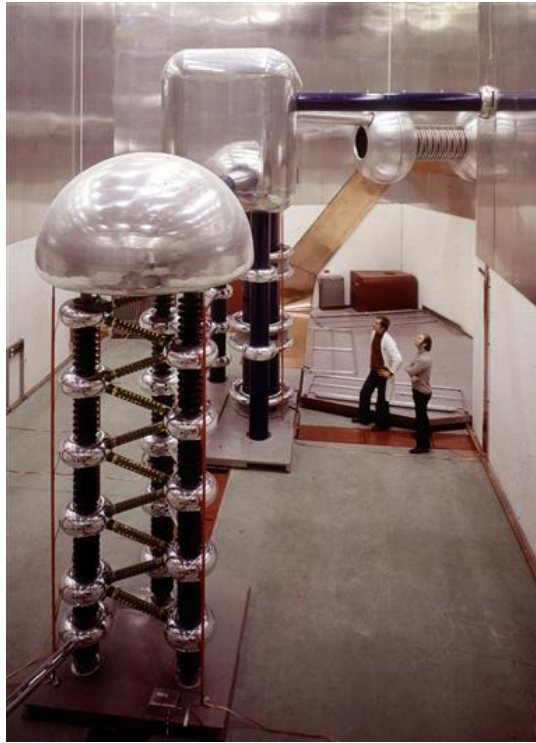

**Fig. 1:** The old (1976–1993) HV installation of CERN Linac2, with the 750 kV Cockcroft–Walton in the front, the ion source electronics in its HV cabin in the background, and the sphere containing the proton source and the HV insulation to the right. The linac is in another room to the right.

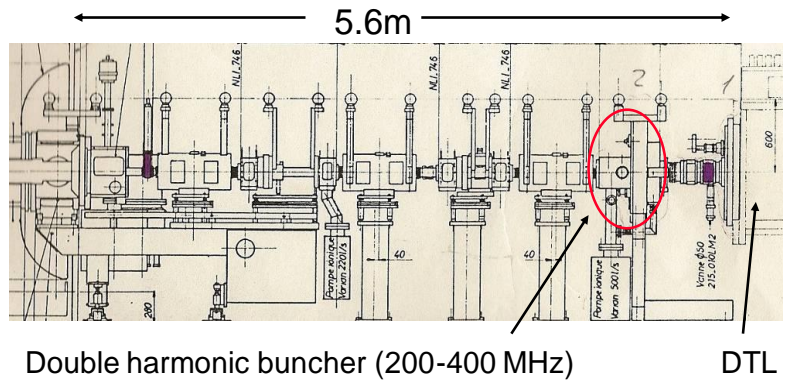

**Fig. 2:** Scheme of the old (1976–1993) 750 keV line of CERN Linac2, between the source of Fig. 1 and the DTL

## 2 The invention of the RFQ

During the 1960s and 1970s the need to build higher current proton accelerators pushed several teams, in particular in the USA and USSR, towards studying solutions to overcome the current limitations of conventional low-energy linac sections. In particular, Ilya Kapchinsky of the Institute for Theoretical and Experimental Physics (ITEP) in Moscow made a significant progress in understanding the behaviour of space charge dominated linac beams and in the frame of his studies started to develop the idea of using at low energy an electric quadrupole focusing channel excited at RF frequency as an alternative to standard electromagnetic quadrupoles. Electric quadrupole forces do not decrease for low particle velocity as the Lorentz force of a magnetic quadrupole field; if the electric field is generated by a RF wave, a beam of particles travelling on the axis of the electric quadrupole will see an alternating gradient resulting in a net focusing force. Kapchinsky's revolutionary idea was to add to the electrodes producing the quadrupole field a longitudinal "modulation" (i.e. a sinus-like profile) which generates an additional longitudinal electric field component. By matching this longitudinal time-varying field with the velocity and phase of the particle beam it was possible to use this structure for bunching and for a moderate acceleration (more on the functioning of the RFQ will be presented in the next section). The problem of generating the quadrupole RF field was not at all trivial, and it was tackled by another Russian scientist, Vladimir Teplyakov of the Institute of High Energy Physics (IHEP) in Protvino. Together, Kapchinsky and Teplyakov published a first paper in 1969 which was the starting point for the development of the RFQ and resulted in the construction of a first experimental device in 1974 [2]. Although not classified, Kapchinsky and Teplyakov papers were published only in Russian and their work was not known in the West until 1977 when a Czech refugee brought a copy of their original paper to the linac team at Los Alamos in the USA and translated it into English. The Kapchinsky and Teplyakov device immediately looked like the long-time sought idea for generating very high currents at low energy. The Los Alamos team immediately embraced this idea and started a programme to improve it and to produce a first technological test. Los Alamos contributions consisted mainly in the development of an input radial matching section to the focusing channel and in a new resonator design that greatly reduced the non-quadrupole field components. A first proof-of-principle (POP) RFQ aimed at fusion material testing was built at Los Alamos and successfully commissioned in 1980 [3]; although it operated only for a few hours before being damaged while increasing the duty cycle, this first operational RFQ demonstrated the validity of the principle and paved the way for the successive developments. During the 1980s, the RFQ design was constantly improved and made more reliable, and RFQs started to replace the HV injectors of the main accelerator laboratories. At CERN, a first RFQ was built already in 1984, and in 1993 a second more powerful RFQ, the 200 mA RFQ2, eventually replaced the Cockcroft–Walton injector and the transport line of Linac2 [4]. Figure 3 shows the new RFQ2 (202 MHz, 1.8 m length) in front of the Linac2 DTL, with its proton source and LEBT. This new compact system occupies the same floor surface as the old 750 keV transport line of Fig. 2; the installation of the RFQ allowed decommissioning the entire HV injector of Fig. 1.

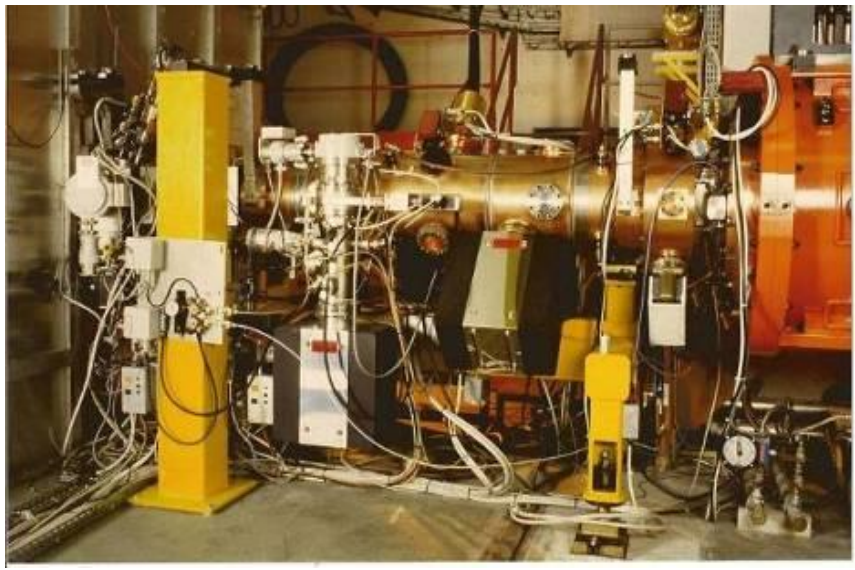

**Fig. 3:** The new 750 keV RFQ2 (1993) installed in front of CERN Linac2

## 3 The three RFQ functions

The reason why the RFQ became so popular is that it fulfils at the same time three different functions:

(i) *focusing* of the particle beam by an electric quadrupole field, particularly valuable at low energy where space charge forces are strong and conventional magnetic quadrupoles are less effective;

(ii) *adiabatic bunching* of the beam: starting from the continuous beam produced by the source it creates with minimum beam loss the bunches at the basic RF frequency that are required for acceleration in the subsequent structures;

(iii) *acceleration* of the beam from the extraction energy of the source to the minimum required for injection into the following structure.

In modern systems the ion source is followed by a short LEBT required to match transversally the beam coming from the source to the acceptance of the RFQ. Extraction from the ion source (and injection into the RFQ) is usually done at an energy of a few tens of kiloelectronvolts, achievable with small size HV installations. The RFQ follows the LEBT, and accelerates the beam up to entrance of the following structure, usually a DTL. Although a RFQ could accelerate the beam to any energy, most of the RF power delivered to the resonator goes to establishing the focusing and bunching field, with the consequence that its acceleration efficiency is very poor. For this reason, RFQs are used only in the low-energy range, up to few megaelectronvolts for protons, and their length usually reaches a maximum of a few metres. As soon as the beam is bunched and the energy is sufficiently high, it is economically convenient to pass to another type of accelerating structure. Figure 4 shows a photograph of the inside of a RFQ (CERN RFQ1, 202 MHz) and a three-dimensional view of the CERN RFQ for Linac4 (352 MHz), presently under construction.

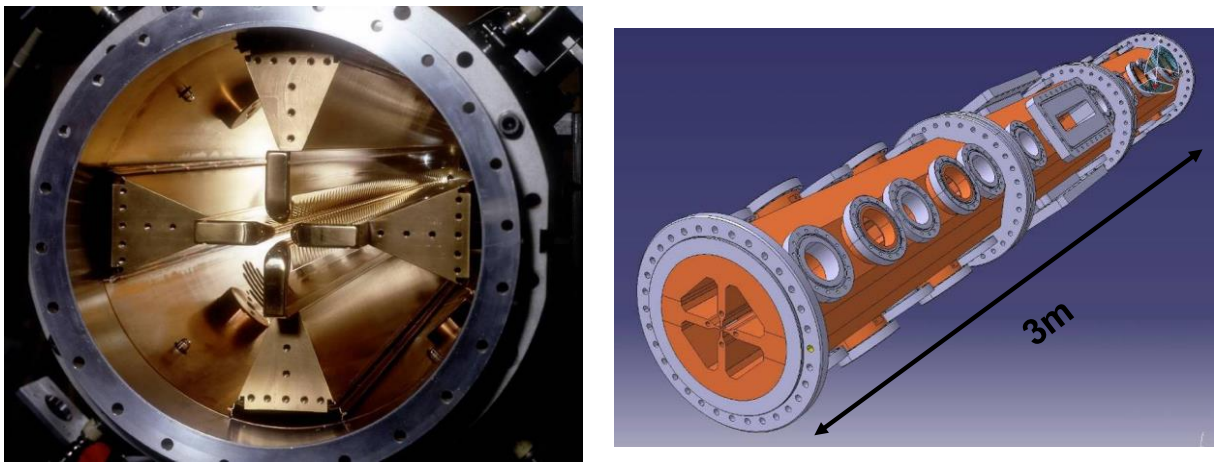

**Fig. 4:** The CERN RFQ1 (left) and Linac4 RFQ (right)

The generation of the quadrupole electric field requires four electrodes, visible in the left-hand side photograph of Fig. 4, which in this particular type of RFQ are called "vanes". They are positioned inside a cylindrical tank forming a RF cavity which resonates in a mode that generates a quadrupole RF voltage between the vane tips (Fig. 5). A particle travelling through the channel formed by the four vanes will see a quadrupole electric field with polarity changing with time, at the period of the RF. Every half RF period the particle will see the polarity of the quadrupole reversed, i.e. it will see an alternating gradient focusing channel, with periodicity corresponding to the distance travelled by the particle during half RF period, i.e. $\beta\lambda/2$. The physics of this electric quadrupole channel is the same as for a magnetic focusing channel where the quadrupole gradient is replaced by the RF voltage and the space periodicity is $\beta\lambda/2$.

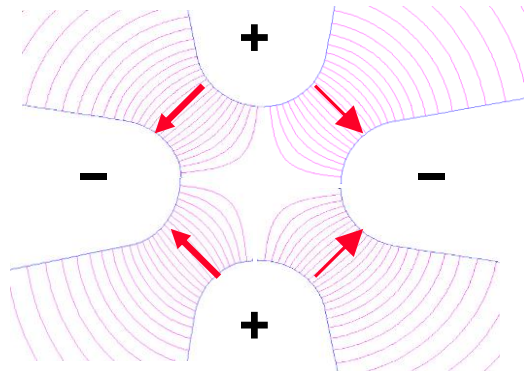

**Fig. 5:** Voltages and electric fields across RFQ vanes

The longitudinal focusing required for bunching and acceleration is provided by a small longitudinal modulation of the vane tips (barely visible in Fig. 4 left). On the tip of the vanes is machined a sinusoidal profile, with period $\beta\lambda$ (Fig. 6). The important point, necessary to obtain a longitudinal field component, is that on opposite vanes peaks and valleys of the modulation correspond, whereas on adjacent (at 90°) vanes peaks correspond to valleys and vice versa (Fig. 6, with adjacent vanes presented on the same plane for convenience). The arrows in the scheme for adjacent vanes of Fig. 6 represent at a given time the electric field between the two adjacent vanes which have opposite polarity (voltage difference *V*). On the axis, the electric field vectors can be decomposed into a transverse component, perpendicular to the direction of the beam, and in a small longitudinal component, parallel to the beam direction. The transverse component is constant along the length and represents the focusing field. The longitudinal component instead changes sign (direction) every $\beta\lambda/2$: a particle travelling with velocity $\beta$ will see an accelerating field (or, in more general terms, the same RF phase) in every cell, exactly as in a standard $\pi$ mode accelerating structure.

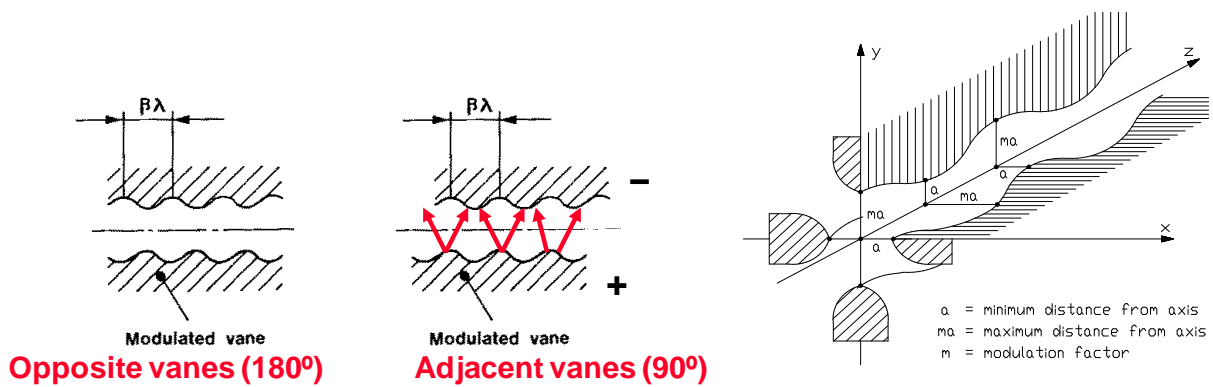

**Fig. 6:** RFQ vanes, field polarity and modulation parameters

As a result, from the longitudinal point of view a RFQ will be made of a large number of small accelerating cells ($\beta$ being very small at the beginning of the acceleration), with the additional flexibility with respect to conventional structures that it is possible to change from cell to cell: (a) the amplitude of the modulation and therefore the intensity of the longitudinal electric field; and (b) the length of the cell and therefore the RF phase seen by the beam in its centre. It is then possible to keep the vanes flat in the initial part of a RFQ (no modulation and only focusing) and after a certain length start ramping up slowly the modulation and the longitudinal field. After the first modulated cells, the particle density will start increasing around the phase at which the RF voltage passes through zero and the bunch will be slowly formed. Over many cells, the bunching process can be carefully controlled and made "adiabatic", with the result of capturing a large fraction of the beam inside the RFQ "bucket". When the bunch is formed, the acceleration can start, and the RFQ designer can slowly modify the cell length to bring the centre of the bunch towards the crest of the RF wave. As an example, Fig. 7 shows the evolution of the longitudinal beam emittance (energy versus phase) in 8 selected cells out of the 126 that make the CERN RFQ2 of Fig. 3 (90 keV to 750 keV): in one of the first modulated cells (top left) the continuous beam coming out of the source sees the first sinusoidal energy modulation; in the following cells (first line) the bunching proceeds until a sufficient density is achieved in the centre (second line) and the acceleration process can start. A few particles are lost in the process, corresponding to the tails visible in the fourth and fifth plots. In the last cells, a bunch is formed, ready for injection into the DTL.

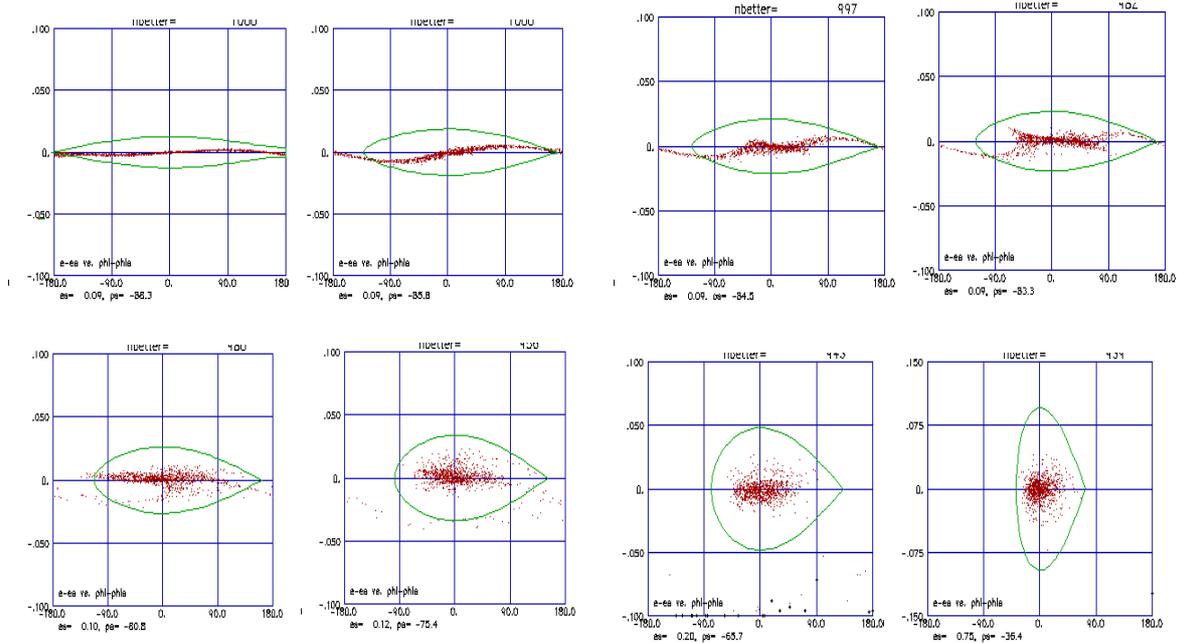

**Fig. 7:** Evolution of longitudinal emittance along the CERN RFQ2, in 8 representative cells out of 126

Again, it must be observed that in a RFQ only the last cells are devoted to acceleration; a RFQ is mainly a focusing and bunching device. By correctly defining the parameters of the modulation and the RF voltage, the beam dynamics designer is able to match and transport intense beams, at the same time bunching the beam with minimal particle loss. The drawback is that the beam focusing parameters are frozen forever in the beam modulation and cannot be changed during operation; the RFQ is a "one-button" machine, where only the RF voltage can be varied during operation. Its design relies completely on the beam transport codes, and it is not by coincidence that the development of the RFQs has gone in parallel with the development of the modern powerful beam simulation codes which are able to correctly treat the space charge regime.

## 4  A brief introduction to RFQ beam dynamics

As seen in the previous section, from the point of view of the beam a RFQ is made of a sequence of hundreds of cells with the simplified shape shown in Fig. 8. The dimensions of the region between the electrodes is small compared with the RF wavelength, thus the electric field between the vanes can be calculated in quasi-static approximation and depends only on the geometry of the electrodes. For each RFQ cell, the beam dynamics designer can use three parameters to define the action of the cell on the beam:

1. the aperture *a*, which defines the focusing strength;

2. the modulation factor *m*, which defines the intensity of the longitudinal field component;
3. the phase *φ*, which is given by the difference between the ideal modulation period (*βλ*/2) and the real one, and defines the bunching and/or accelerating action.

These parameters are specific to each cell, and can be changed, although smoothly, between one cell and the next. On top of them, the designer can act on another parameter that is common to all cells (or can be changed in more sophisticated designs, but with limited freedom), the RF voltage *V*.

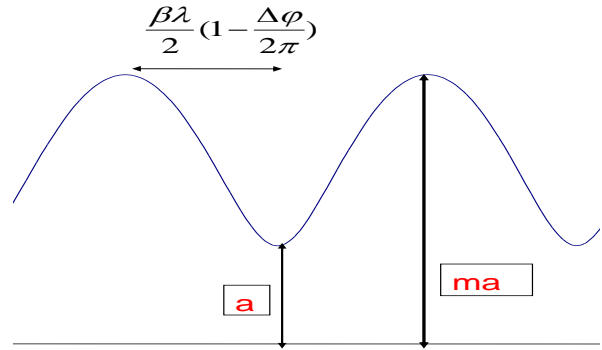

**Fig. 8:** Parameters of a RFQ cell

Designing a RFQ modulation consists of finding an appropriate set of (*a, m, ϕ*)$_i$ for each cell *i* together with a voltage *V* which allows realizing in the minimum possible number of cells the following functions:

- matching of the beam out of the LEBT into the RFQ focusing channel;
- transporting the beam (transversally) with minimum emittance growth;
- bunching the beam with minimum beam loss, generating a longitudinal emittance matched to the acceptance of the following accelerator;
- accelerating the beam from the source extraction energy up to the energy required for injection into the following accelerator;
- for some more modern designs, matching the beam to the following structure using the last cells of the RFQ.

This design is usually done by computer codes. Several programs exist that from a set of input parameters and for a given voltage define the (*a, m, ϕ*) sequence and calculate the output beam parameters, the first and most famous being PARMTEQ, developed at Los Alamos for the POP RFQ [5]. The experience of the designer remains particularly important, however, in determining the impact of a given set of parameters on the other aspects of the RFQ, its RF and mechanical design and construction. In particular, large voltages increase the focusing but increase as well the risk of voltage breakdowns between the electrodes, too small apertures could lead to unrealistic tolerances in the electrode machining and alignment, etc. Moreover, the designer needs to devote particular care in simulating the beam evolution in presence of realistic error distributions, in particular on the positioning of the electrodes. Often the best design is not the one that gives the best performance (short RFQ, small emittance growth, small beam loss) but the one that is less sensitive to mechanical and RF errors.

Before analysing how the modulation parameters translate into beam dynamics parameters, it is important to consider how the two-dimensional treatment considered so far translates into the fully three-dimensional shape of a real electrode. In particular, we consider the vanes of a "four-vane" RFQ (Fig. 4). The original approach developed by Kapchinsky was purely analytical, at a time when powerful computer codes were not available, and allows a good insight of the RFQ field. The starting assumption is that in the static approximation that we are allowed to use the potential must be a solution of the Laplace equation, which in cylindrical coordinates can be represented by a series of Bessel functions. The basic Kapchinsky's idea was that of all of the terms in the series of Bessel functions only two were required for a focusing and accelerating beam channel: a transverse quadrupole term and a longitudinal sinusoidal term. In mathematical form, this means that the voltage in cylindrical coordinates has to be written as the sum of the two Bessel components:

$$V(r, \vartheta, z) = A_0 r^2 \cos 2\theta + A_{10} I_0(kr) \cos kz \tag{1}$$

with $k=2\pi/\beta\lambda$. The voltage on the surface of the metallic electrodes must be constant, and this means that the three-dimensional profile of a RFQ electrode must correspond to an equipotential surface of $V(r, \theta, z)$. Such surfaces are hyperbolae in the transverse plane, presenting longitudinally the characteristic sinusoidal modulation. The electrode profile is still defined by the parameters of Fig. 7; a detailed mathematical analysis shows that the constants $A_0$ and $A_{10}$ can be expressed in terms of the modulation parameters and of modified Bessel functions as

$$A_0 = \frac{V_0}{2a^2} \frac{I_0(ka) + I_0(kma)}{m^2 I_0(ka) + I_0(kma)} \qquad A_{10} = \frac{V_0}{2} \frac{m^2 - 1}{m^2 I_0(ka) + I_0(kma)} \tag{2}$$

To provide a pure quadrupole field the transverse faces of the four electrodes have to follow a hyperbolic shape; however, the electrode cannot extend indefinitely, and the hyperbola has to be truncated at a certain position. In this respect, different configurations are possible. In early RFQ designs as the RFQ1 of Fig. 7, the transverse profile followed precisely the hyperbolic shape up to a few centimetres from the vane tip, introducing only a small negligible deviation from the pure quadrupole potential. In later designs, the mechanical construction has been greatly simplified by either taking a circular cross-section for the vane tips or by even taking as electrode a circular rod instead of a vane. Such mechanical simplifications introduce multipole components that can be calculated by computer codes and whose effect on the beam can be minimized.

A complete treatment of the RFQ beam dynamics can be found in several books and reports [5–7]; here, to understand the main features of the RFQ design it is important to give the main relations that connect the RFQ design parameters (*a, m, ϕ*) and *V* with the conventional beam dynamics parameters used in a linear focusing and accelerating channel, transverse focusing coefficient *B* and longitudinal field $E_0T$:

$$B = \left(\frac{q}{m_0}\right)\left(\frac{V}{a}\right)\left(\frac{1}{f^2}\right)\frac{1}{a}\left(\frac{I_o(ka) + I_o(mka)}{m^2 I_o(ka) + I_o(mka)}\right) \tag{3}$$

$$E_0 T = \frac{m^2 - 1}{m^2 I_o(ka) + I_o(mka)} \cdot V \frac{2}{\beta \cdot \lambda} \frac{\pi}{4} \tag{4}$$

An example of RFQ beam dynamics design is presented in Fig. 9; here are shown the profiles of *(a, m, ϕ)* along the length of the new Linac4 RFQ at CERN [8].

The Linac4 RFQ operates at 352 MHz frequency, accelerating a 70 mA beam from 45 keV up to 3 MeV energy. It is made of 303 cells for a total length of 3 m. The RF phase seen by the beam at the entrance of the RFQ is –90° (in the linac convention, counted from the crest of the wave), but soon as the bunching process starts the phase is slowly increased to reach –30° after about 120 cm. At that point the modulation factor which is very small at the beginning can be increased, ramping up the longitudinal field and starting the actual acceleration process, which takes place in the second half of the RFQ. Owing to the requirements on the length of this particular RFQ, which could not exceed 3 m for manufacturing reasons, the bunching process takes places relatively quickly and the design beam transmission is only 95%. A theoretical transmission close to 100% is possible, but at the cost of having a longer RFQ with tight mechanical requirements.

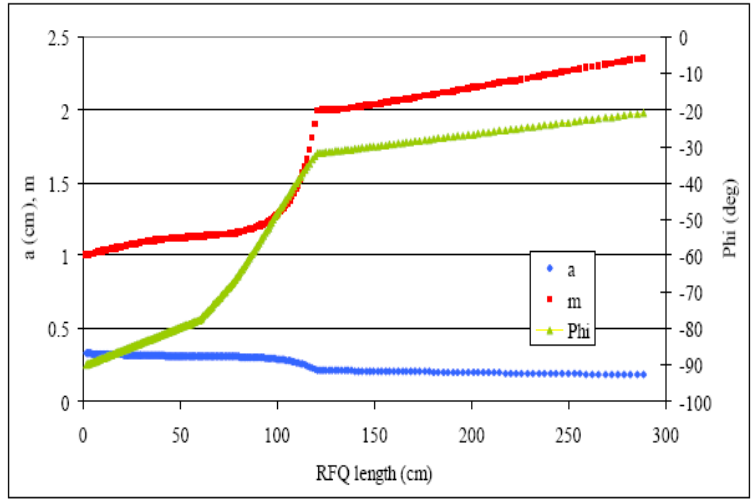

**Fig. 9:** Modulation parameters along the CERN Linac4 RFQ

## 5    The RFQ RF resonator

From the RF point of view, the problem of building a RFQ consists in creating a time-varying quadrupole-type electric field between four electrodes, keeping the voltage constant (or following a pre-defined law) along its length. To generate this field, the electrodes must be part of a RF resonator; different resonator types can be used, the most commonly used being the "four-vane" resonator, developed at Los Alamos for the POP RFQ. It can be considered as a cylindrical resonator where is excited the TE210 mode, i.e. a quadrupole mode (mode index 2 in the angular polar coordinate) with only transverse electric field components and constant fields along its length (mode index 0 longitudinally). The TE210 mode of the empty cylinder, whose electric and magnetic field symmetry is shown on the left side of Fig. 10, is transversally "loaded" by the four vanes that concentrate the electric field on the axis (Fig. 9, right). The RFQ will result in cylinder containing the four vanes, which must be connected to the cylinder all along their length.

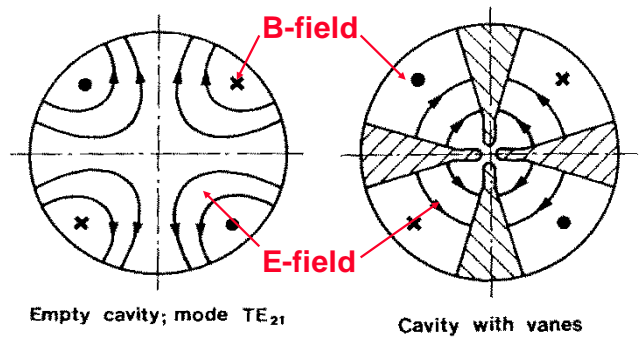

**Fig. 10:** Four-vane RFQ

The vanes have a twofold effect on the TE210 mode: on the one hand, they concentrate the quadrupole field on the axis, increasing the RF power efficiency of the structure and the focusing term $V/a$ in Eq. (3); and, on the other hand, they increase the capacitance for this particular mode, decreasing its frequency well below that of the many other modes of the cylindrical resonator; this separation has a positive effect on the stability of the resonator. Unfortunately, the presence of the vanes decreases in the same way as the frequency of the TE110 mode, the dipole whose field pattern is shown in Fig. 11: the RFQ resonator will present at a frequency slightly below that of the operating TE210 mode two dipole modes of TE110 type, corresponding to the two orthogonal polarizations of this mode.

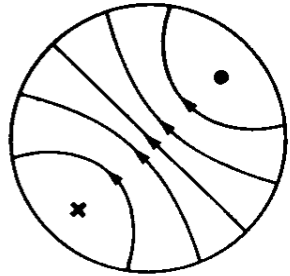

**Fig. 11:** Dipole modes in a cylindrical cavity

After having connected the vanes to the cylinder, it is important to terminate correctly the resonator at its two ends. Longitudinally, the voltage between the vanes must be constant, meaning that the mode of operation must be a pure TE210. The problem is that in normal conditions the TE210 is forbidden in a closed cylindrical resonator: its electric field is directed transversally to the axis, and on the end discs closing the resonator the electric field would be parallel to a metallic wall. To allow the excitation of this mode, the two end regions of the RFQ must be modified, by cutting an opening at the end of each vane (the vane "undercuts") as shown in Fig. 12. The undercuts allow the magnetic field which goes longitudinally in each quadrant to turn around the vanes and continue in the next quadrants. The ends of the vanes do not touch the covers, but leave a small gap where the turning magnetic field excites an electric field. If the resulting "end cell" is made resonant at the frequency of the TE210 mode, the electromagnetic field of the mode will see an infinitely long RFQ (i.e. will not see the presence of the end cells) and the voltage along the vanes will be constant. It should be mentioned that the correct design of the RFQ end cells requires an extensive use of three-dimensional RF simulation codes.

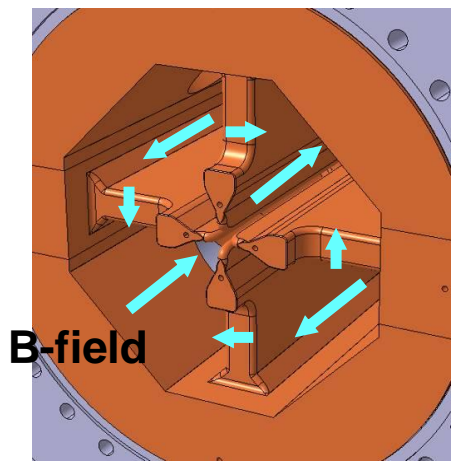

**Fig. 12**: End-cell of a four-vane RFQ. The arrows show the direction of the magnetic field.

The length of the RFQ, as defined by the beam optics, has an important impact on the number and distribution of high-order modes in the RFQ cavity, which will eventually determine the sensitivity of the RFQ voltage to errors in the positioning of the vanes. In a cylindrical resonator each zero-order mode, such as the TE210 and TE110, gives rise to a family of high-order modes of increasing frequency, each one characterized by a longitudinal voltage distribution with a number of nodes (transitions through zero) equal to the order of the mode. The presence of the vanes will lower the frequency of all of the modes of the TE21$n$ and TE11$n$ families, bringing them in a frequency range close to the operating frequency. In particular, the distance between each high-order mode and its zero mode will be inversely proportional to $(l/\lambda)^2$, the square of the ratio between the RFQ length and the RF wavelength of the zero mode, as can be easily derived from waveguide theory. The consequence is that the longer the RFQ the lower will be the spacing between the operating mode and the higher-order modes, opening the possibility of harmful effects on the field stability of the resonator. Although the RFQ operates at fixed frequency on the TE210 mode, the presence of mechanical errors in the machining and/or positioning of the vanes will give rise to field components of the adjacent modes appearing at the operating mode frequency, whose amplitude will be proportional to the mechanical error and inversely proportional to the difference in frequency between operating and perturbing mode. The consequence is that the longer the RFQ, the more stringent will be the mechanical tolerances.

To keep the construction tolerances at a reasonable level, RFQs that are longer than about $2\lambda$ are usually equipped with special compensation devices, e.g. tuning volumes inserted inside the quadrants at different longitudinal positions allow the mechanical errors on the vanes to be compensated for by a local variation of the quadrant inductance. For RFQs that are even longer, from about $4\lambda$, the local compensation is not sufficient and a stabilization scheme is usually implemented, under the form of a resonant or non-resonant device mounted inside the RFQ which moves the frequency of the perturbing modes away from the operating mode. As an example of a long RFQ using only compensation schemes, Fig. 13 shows the measured mode spectrum of a 425 MHz four-vane RFQ, 2.75 m long. For this RFQ, $l/\lambda=3.9$: the zero quadrupole (TE210) is surrounded by a large number of modes, with as many as three dipole modes (TE110, TE111 and TE112) at frequencies lower than the operating frequency. Each dipole mode has two polarizations, corresponding to orthogonal orientations of the electric field (see Fig. 11). These can have slightly different frequencies, each one generating its own high-order band; the notations 1-3 and 3-4 in Fig. 13 refer to the polarizations corresponding to field concentrated in pairs of opposite quadrants.

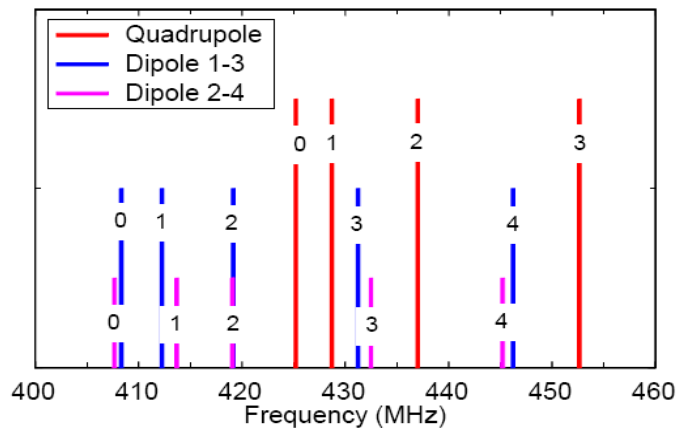

**Fig. 13:** Measured mode spectrum of a 425 MHz four-vane RFQ, 2.75 m long. The notations 1-3 and 3-4 refer to pairs of opposite quadrants. Excitation was in quadrant 1. Operating mode is quadrupole 0.

The high sensitivity to errors of the RFQ resonator coming from the presence of perturbing modes has to be correctly taken into account in the design, construction and tuning of the RFQ. As a first step, the length of the RFQ and the design of the end terminations have to be chosen in such a way as to avoid having dipole modes too close to the operating quadrupole mode. In the case of a RFQ with compensation scheme, as are most of the existing RFQs, an extensive series of voltage measurements is required after construction and assembly. The measurements are then entered into an algorithm that allows the correct dimensioning of the compensation devices: this is the so-called "tuning" of the RFQ, which on top of bringing the quadrupole frequency at the required design value aims at achieving a flat (or following a predefined law) voltage along the RFQ, equal in the four quadrants. Accurate field measurements in the RFQ can be performed via "bead-pull" techniques, where a perturbing metallic bead on a plastic wire is slowly moved inside the four quadrants, to register the frequency shift which is proportional to the square of the local field.

To reduce the error sensitivity of the RFQ field, resonator designs alternative to the four-vane have been developed and are in use in many laboratories; of these, the most widely used is the so-called "four-rod" RFQ (Fig. 14) originally developed by A. Schempp at the IAP of Frankfurt University [9].

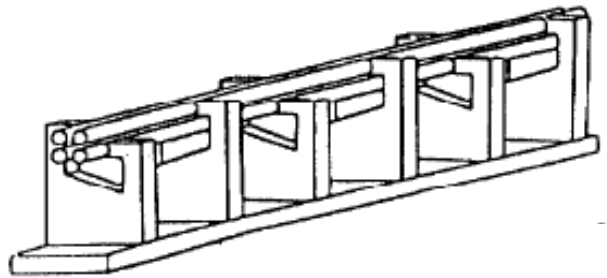

**Fig. 14:** Four-rod RFQ

In this device, the four electrodes are either circular rods with a modulated diameter or small rectangular bars with a modulated profile on one side; they are connected to an array of quarter-wavelength parallel plate transmission lines generating a voltage difference between the two plates (Fig. 15). Opposite pairs of electrodes are connected to the two plates of a line, resulting in a quadrupole voltage being generated between the rods. Several quarter-wavelength cells are used to cover the required RFQ length; their magnetic field couples from each cell to the next, forming a single long resonator. This "open" RFQ structure is then placed inside a tank, which forms the vacuum and RF envelope of the structure.

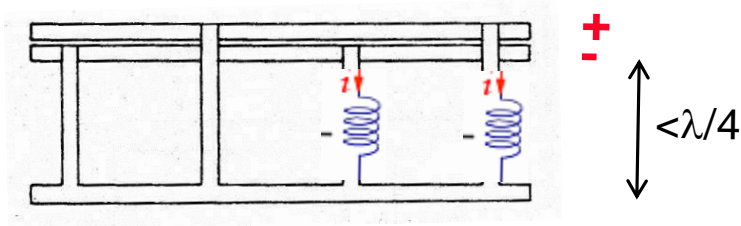

**Fig. 15:** The four-rod RFQ structure

RFQs of this type are free from dipole modes; however, in the standard single-support structure of Fig. 14 the transverse slope in the cut on the plate needs to be carefully defined to ensure that the voltage on two pairs of opposite rods is the same, to compensate for their different distances from the bottom plate. In other RFQ designs, such as that shown in Fig. 16, a double plate supports the electrodes, to completely eliminate dipole components. From the point of view of the longitudinal modes a four-rod RFQ is no different to a four-vane RFQ: for long RFQ structures an error compensation is required, achieved by placing short-circuiting plates or metallic volumes inside some cells to reduce the transmission line length.

The main advantages of the four-rod RFQ are the absence of dipole modes (which reduces the sensitivity to mechanical errors and simplifies the tuning), the reduced transverse dimensions as compared with the four-vane RFQ, and the simple and easy to access construction. These advantages are particularly evident for the low-frequency RFQs (up to about 100 MHz) used for heavy ions. For the higher frequencies required for protons, from about 200 MHz, the transverse dimensions of the four-rod RFQ become very small and the current and power densities reach high values in some parts of the resonator, in particular at the critical connection between the rods and the supports; cooling can be difficult, in particular for RFQs operating at high duty cycle, with the risk of excessive deformations of the rods and reduced beam transmission. For these reasons, RFQs operating at frequencies above 200 MHz or at high duty cycle are usually of the four-vane type.

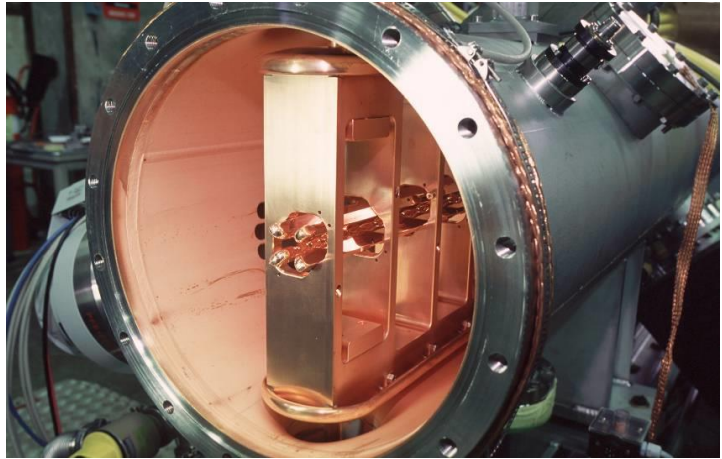

**Fig. 16:** The antiproton decelerating RFQ at CERN (202 MHz)

## 6    Mechanical construction

The mechanical design and construction of a RFQ is a challenge in itself, for two main reasons: first of all, tight tolerances in the machining and positioning of the electrodes need to be achieved and maintained during operation; and, second, because many different mechanical parts need to be joined together respecting the tolerances and providing an excellent electrical and thermal contact, to avoid excessive RF power consumption and/or overheating. On top of that, the mechanical structure has to provide sufficient access points for RF tuning and for vacuum pumping.

Usual beam dynamics tolerances in the machining and positioning of the RFQ electrodes are of the order of few tens of micrometres, a value that corresponds to about 1 % of the minimum radius of the beam channel ($a$ in Fig. 7). For larger errors, multipoles appear in Eq. (1) resulting in a perturbation of the beam optics and in increasing beam loss in the RFQ. In four-vane RFQs the RF can generate additional dipole and higher-order mode components again proportional to the errors in the positioning of the vanes through complex RF-related algorithms. RF-related errors can be compensated for by the compensation system (tuners or other); nevertheless, the maximum permitted vane positioning errors define the size of the compensation system (number and dimension of the tuners, for example). The compromise usually adopted in this type of RFQs is to define for the RF a maximum error smaller or of the same level as the beam dynamics one, and then dimension consequently the compensation system. To give an example, Table 1 reports the permitted error budget of the CERN Linac4 RFQ, defined after a series of beam dynamics calculations in presence of random errors. The RF compensation system (tuners) is dimensioned to fully absorb the mechanical errors, leaving a residual field error of ±1 %. The electrode gap represents the distance between the vane ends at the connection between the three RFQ segments.

Table 1: Error budget of the CERN Linac4 RFQ

| Linac4 RFQ Mechanical Tolerances | Value | Units |
|---|---|---|
| Machining error | ± 20 | μm |
| Vane modulation error | ± 20 | μm |
| Vane tilt over 1 m | ± 100 | μm |
| Vane positioning error (displacement h+v) | ± 30 | μm |
| Vane thickness error | ± 10 | μm |
| Gap between vanes (contiguous modules) | 100 ± 15 | μm |
| Section tilt over 1 m | ± 30 | μm |
| Electromagnetic field error | ± 1 | % |

Joining the different parts is the next problem to be faced; RFQs tend to use a large variety of joining techniques, including brazing, electron-beam welding, TIG welding and simple bolting of parts using different types of gaskets and contacts. Four-rod RFQs are usually made out of parts bolted together; low- and medium-frequency four-vane RFQs are bolted or welded, and high-frequency four-vane RFQs are usually made of brazed copper elements following the scheme of Fig. 17. The RFQ is divided into longitudinal segments of about 1 m length (Fig. 4, right); the segments are composed of four copper elements brazed together, each made of a vane and of a part of the external tank to minimize the brazing surface. Cooling channels are machined inside the copper; the brazing ensures the vacuum tightness of the structure, providing at the same time a high thermal and electrical conductivity.

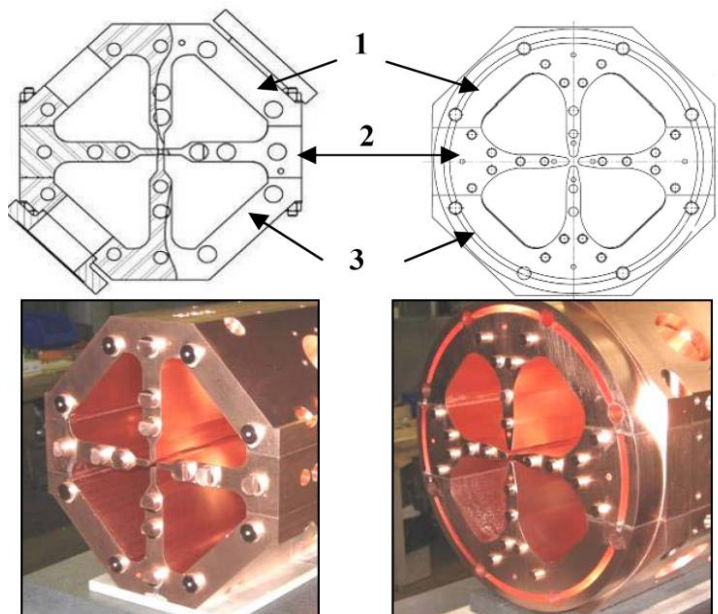

**Fig. 17:** Construction scheme of two European 352 MHz continuous wave RFQs: TRASCO of INFN, Italy [10] (left) and IPHI of CEA, France [11] (right)

An important requirement is that the precise alignment of the electrodes does not change when the structure is heated by the RF power dissipated on the walls and supports; this is particularly demanding for high duty cycle or continuous wave RFQs, where the power to dissipate can be of the order of 1 kW/cm. The number size and position of the water cooling channels need to be carefully dimensioned; the corresponding deformations have to be calculated and translated into beam dynamics and RF errors. In addition, a precise control of the water temperature is required, at the level of 0.1°. As an example, Fig. 18 summarizes the thermal studies performed for the design of the TRASCO RFQ of INFN. Here the deformations are translated into frequency errors in the individual RFQ quadrants.

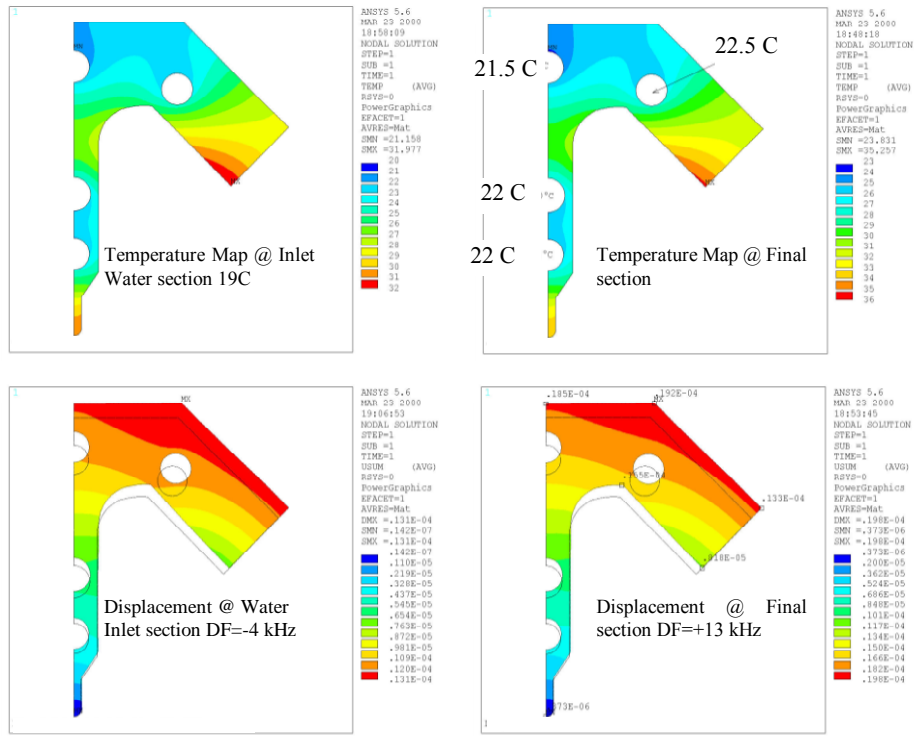

**Fig. 18:** Calculated temperature (top) and displacement (bottom) distributions in the TRASCO RFQ at the beginning (left) and at the end (right) of a section [10]

# 7   Putting it all together

In the previous sections the main aspects of RFQ theory and construction practice have been presented. What remains to be underlined now is that a RFQ is a highly multidisciplinary object, whose performance relies on a complex equilibrium between three fundamental disciplines, beam dynamics, electrodynamics and accelerator mechanics, and with important inputs from other aspects of accelerator technology such as vacuum, RF power production, survey, etc. On the one hand, mechanical errors or deformations and deviations from the ideal RF field distribution would result immediately in reductions of the RFQ beam transmission; on the other hand, an excessively demanding tolerance budget would increase the complexity of the construction leading to unnecessary challenges and sky-scraping costs. On top of that, the performance of a RFQ depends critically on its input beam parameters and therefore on the performance of the ion source and LEBT: any deviation from the design emittance or error in the input beam alignment result again in a reduction of the RFQ transmission.

The real challenge of building a RFQ is not in each single aspects of its design, but lies in putting it all together: if teamwork and good communication between the different competencies required to build an accelerator are nowadays crucial to any project, this is even more true for RFQs where the different aspects are closely interrelated. The fact that often projects are based on international or inter-laboratory collaborations and/or rely heavily on industrial partners adds another degree of complexity that needs to be correctly managed.

If the construction of a RFQ represents a challenge, it can also be extremely rewarding. RFQs in particular for extreme parameters tend to be difficult to design and construct, but once they have been commissioned they tend to be very reliable (as far as their thermal equilibrium is not altered), operating steadily and without need for adjustment for several years: after all, they are "one-button" accelerators.

## Acknowledgements

The preparation of this lecture has profited from the support and advice of A.M. Lombardi, A. Pisent, C. Rossi and J. Stovall. To all of them goes my gratitude.